# Managing the Impact of Sensor Thermal Noise in Machine Learning for Nuclear Applications

**Issam Hammad[1][2]**

[1]Department of Engineering Mathematics, Dalhousie University, Halifax, NS, Canada

[2]Engineering Department, Alithya Digital Technology Corporation, Pickering, Ontario, Canada

Issam.hammad@dal.ca

**Abstract**

Sensors such as accelerometers, magnetometers, and gyroscopes are frequently utilized to perform measurements in nuclear power plants. For example, accelerometers are used for vibration monitoring of critical systems. With the recent rise of machine learning, data captured from such sensors can be used to build machine learning models for predictive maintenance and automation. However, these sensors are known to have thermal noise that can affect the sensor's accuracy. Thermal noise differs between sensors in terms of signal-to-noise ratio (SNR). This thermal noise will cause an accuracy drop in sensor-fusion-based machine learning models when deployed in production. This paper lists some applications for Canada Deuterium Uranium (CANDU) reactors where such sensors are used and therefore can be impacted by the thermal noise issue if machine learning is utilized. A list of recommendations to help mitigate the issue when building future machine learning models for nuclear applications based on sensor fusion is provided. Additionally, this paper demonstrates that machine learning algorithms can be impacted differently by the issue, therefore selecting a more resilient model can help in mitigating it.

1. Introduction

The recent rise in computational power has led to remarkable innovations in the areas of artificial intelligence and machine learning. Utilizing machine learning for predictive maintenance, inspection automation, autonomous control, and verification is booming. Many previously published research papers have proposed using machine learning to support the operation and inspection of Canada Deuterium Uranium (CANDU) reactors. An example can be seen in [1], where machine learning is used in fault prediction for the primary heat transport system of CANDU type pressurized heavy water reactors. In [2] machine learning techniques were proposed for the verification of refueling activities in CANDU-type nuclear power plants with direct applications in nuclear safeguards. In [3], deep learning was used to automate the detection of flaws in CANDU nuclear fuel channel ultrasonic testing (UT) scans.

Many machine learning applications for CANDU are built using data resulting from sensor fusion. This means that the model relies on measurements from several sensors to make a specific prediction. Commonly used sensors in CANDU operation include accelerometers which are devices that measure vibration or acceleration of motion, magnetometers which are devices that measure the strength of magnetic fields and often the direction, and gyroscopes which are devices that measure or maintain angular velocity. The utilization of accelerometers, magnetometers, and gyroscopes in the operation and maintenance of CANDU reactors have been discussed in many applications as presented in [4-10]. Machine learning models, in general, are built by splitting the available dataset into training data and cross-validation/test data. In some instances, a test dataset is used separately from the cross-validation one. In the vast majority of machine learning literature research, the cross-validation or test dataset accuracy is reported as the final accuracy of the proposed model. Nevertheless, when dealing with systems that will be operational in production, considerations beyond the achieved test dataset should be taken into account. This was discussed in detail in a previous research paper that I published [11].

One of the most critical issues that should be evaluated when dealing with sensor-based machine learning models is the issue of thermal noise. The impact of thermal noise on the accuracy of the final model in production should be evaluated. This can be a deal-breaker, particularly when a minimum accuracy or sensitivity is required to adopt a specific machine learning solution for CANDU reactors. Thermal noise results in an added accuracy loss in sensor-based machine learning models which depends on the signal-to-noise ratio (SNR). Additionally, the topology and the selected machine learning algorithm impact the tolerance to thermal noise. This paper discusses this issue and presents recommendations to manage it.

This paper is structured as follows; Section 2 discussed the issue of thermal noise in sensors. Section 3 presents various CANDU applications where sensors including accelerometers, magnetometers, and gyroscopes are used. Section 4 discusses how thermal noise can impact machine learning model performance in production. Section 5 presents a list of recommendations to help mitigate the issue. Section 6 presents the research conclusion.

## 2. Thermal Noise

Thermal noise is a known issue that affects electrical circuits. It is also known as Johnson–Nyquist noise. It is caused by the random motion of electrons in conductors due to thermal agitation. Sensors including accelerometers, magnetometers, and gyroscopes all have their own independent thermal noise. This means that even if we have two sensors with the same part number, thermal noise will be unique to each. Thermal noise is white noise and is described as an additive zero-mean gaussian noise. In the literature, various levels of reported thermal noise can be found which vary in terms of SNR in ranges between 0dB-40dB [12-13]. This noise has an impact on the sensor's reading accuracy, therefore it impacts the trained machine learning model inference accuracy. Figure 1 illustrates the high-level steps used in machine learning models which are based on a sensor fusion.

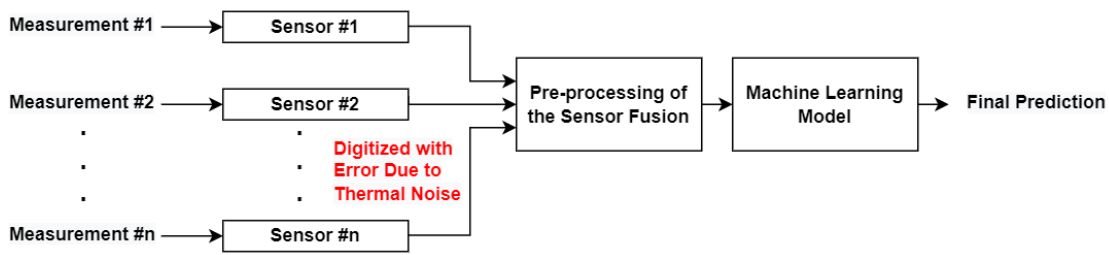

**Figure 1: High-Level Steps used in Machine Learning Models Based on Sensor Fusion**

### 3. Reported Usage of Accelerometers, Magnetometers, and Gyroscope for CANDU

The usage of sensors including accelerometers, magnetometers, and gyroscopes can be found in various CANDU systems. Table 1 lists some of these applications and their reference. Developing a machine learning solution for any relevant application will be susceptible to the thermal noise issue which is discussed in this paper.

**Table 1: Reported Usage of Accelerometers, Magnetometers, or Gyroscopes for CANDU Reactors Operation and Inspection**

| Application | Reference |
| --- | --- |
| Measuring the dynamic response of a bundle of steam generator U-tubes with an Anti-Vibration Bar (AVB) | [4] |
| Investigation of effects of angular misalignment on flow-induced vibration of simulated CANDU fuel bundles | [5] |
| Investigating the consequences of fuel channel failure in a CANDU reactor under normal operating conditions | [6] |
| Modal parameter identification of a CANDU reactor building using ambient vibration measurements | [7] |
| Seismic analysis of a rotor-bearing system | [8] |
| Response of rotating machinery subjected to seismic excitation | [9] |
| Assessment of operability and structural integrity of a vertical pump for extreme loads | [10] |

### 4. Impact of Sensor's Thermal Noise on the Accuracy of Machine Learning Models

To present the possible impact of various levels of sensors' thermal noise on any future machine learning solutions for CANDU, a similar application in biomedical engineering can be utilized. I have published a previous study [11] that discusses the issue and its impact on machine learning models' accuracy based on sensor fusion for a biomedical application.

The chosen application [11] presents a classification problem for 19 physical activities performed by 8 different participants. The data was captured using five 3-DOF (degrees of freedom) orientation trackers which were placed on the participants' bodies. Each orientation tracker includes a 3D accelerometer, a 3D magnetometer, and a 3D gyroscope. Therefore, each orientation tracker provides a total of 9 readings. The dataset was published by [14] and was captured at a frequency of 25Hz. Therefore, every 25 readings represent 1 second in time. Examples of the activities include sitting, standing, jumping, moving around in an elevator, ascending and descending stairs. Several previous papers have used this dataset to propose machine learning models for the classification of physical activities [15-17]. To establish a baseline of accuracies, Table 2 presents the various accuracies that different machine learning models can achieve in classifying this problem. All accuracies were obtained using k-fold cross-validation with k=5.

**Table 2: Reported Accuracies of Various Machine Learning Models**

| Model Description | Achieved Accuracy |
|---|---|
| Deep Neural Network (DNN) | 99.21% |
| Decision Tree Classifier (DTC) following Principal Component Analysis (PCA) | 90.98% |
| Random Forest Classifier (RFC) | 98.74% |
| K-Nearest Neighbors (KNN) following Principal Component Analysis (PCA) | 98.02% |
| Gaussian Naïve Bayes (GNB) | 93.24% |

To simulate the impact of thermal noise on the accuracy of machine learning models, a simulation for various levels of SNR can be performed on the test data. This reflects the reality of this problem that will occur in production. This can be done by generating a noise signal and adding it to the original sensed input to simulate a specific SNR value. Figure 2 illustrates the histogram of one generated noise signal which can result in an SNR value of 5dB. This method of error modeling was adopted in other machine learning research papers such as [18-23]. Figure 3 illustrates an example of the change in the input signal when simulating the impact of a 5dB thermal noise on one accelerometer axis.

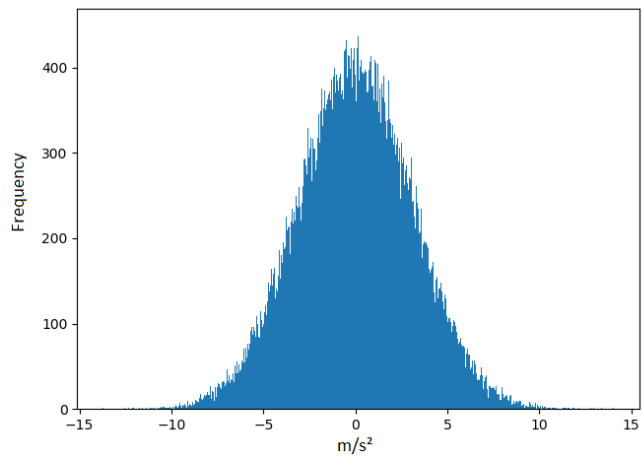

**Figure 2: Histogram for a Generated Noise Signal (5dB SNR Example)**

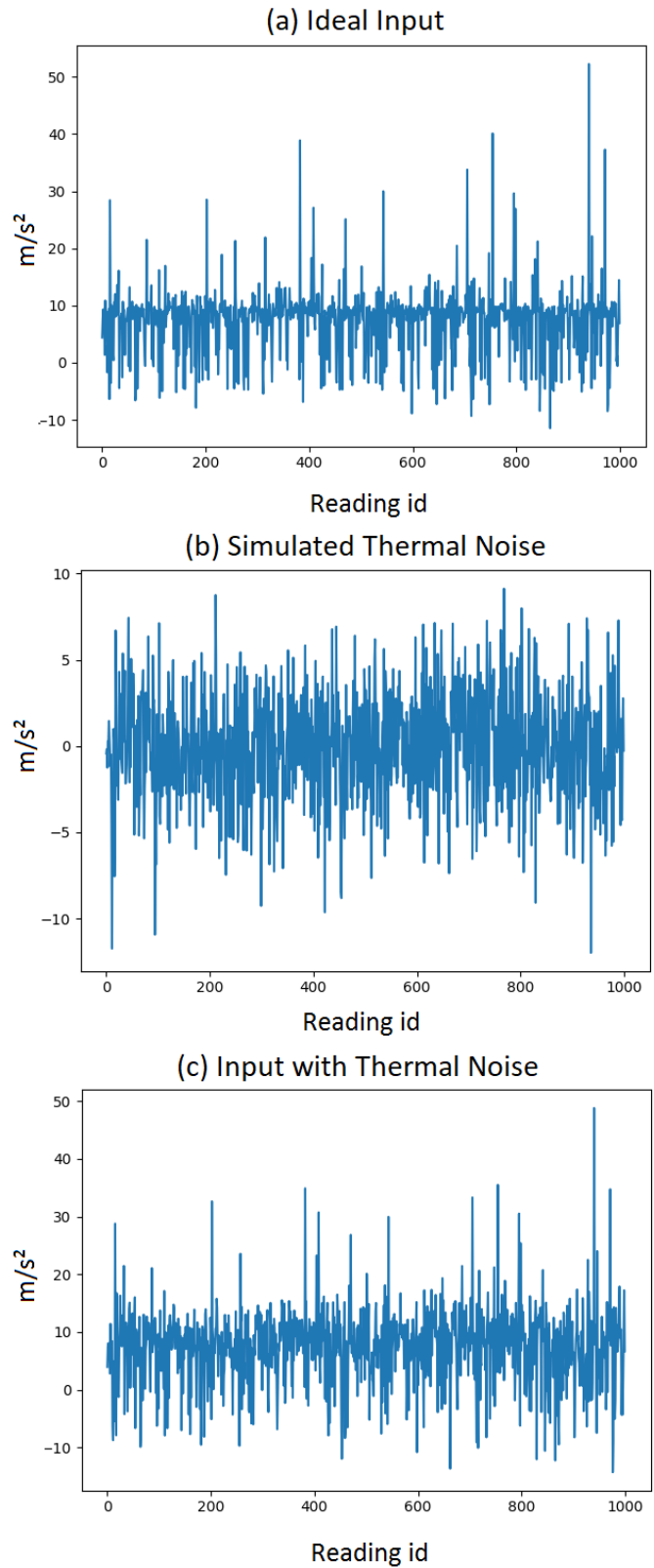

**Figure 3: Simulating the Impact of Thermal Noise (Example of 5dB SNR)**

Table 3 illustrates the accuracy loss resulting from various levels of thermal noise for different machine learning algorithms. As can be seen from the table; algorithms vary significantly in terms of their resilience to noise. Some models can experience significant accuracy loss with a higher level of thermal noise. Therefore, it is fundamentally important to select a model that can be resilient to this issue even if it doesn't have the highest accuracy during development. Different model topologies for the same algorithm can also result in varying results. The table can be seen only as an example of how models can significantly vary in terms of accuracy loss. Therefore, the ideal cross-validation/test accuracy obtained during development doesn't reflect how the model will operate in the field. In summary, a trade-off should be established between the models' accuracy during development and their resilience to noise when selecting a production model.

Table 3: Accuracy Drop Due to Various Thermal Noise Levels

| Algorithm | Thermal Noise SNR and Resulting Accuracy Loss | | | | |
|---|---|---|---|---|---|
| | **40dB** | **30dB** | **20dB** | **10dB** | **5dB** |
| DNN | -0.01% | -0.03% | -0.04% | -0.09% | -0.17% |
| DTC + PCA | -0.14% | -0.89% | -2.43% | -9.4% | -16.1% |
| RFC | -0.04% | -0.29% | -4.03% | -12.06% | -28.3% |
| KNN + PCA | -0.05% | -0.12% | -0.14% | -0.31% | -0.45% |
| GNB | -0.16% | -0.32% | -21.2% | -23.4% | -42.25% |

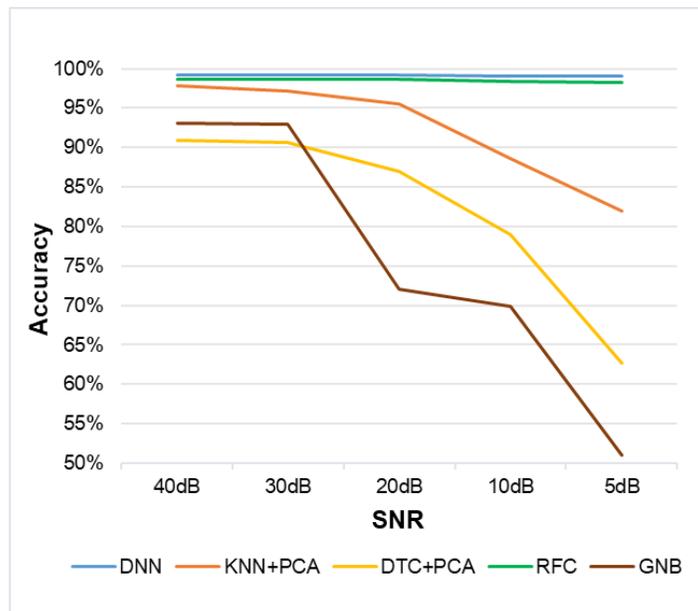

**Figure 4: Accuracy Loss Trend for Various Machine Learning Models Based on SNR**

## 5. Recommendations

Based on the presented problem in this paper, the following list of recommendations can be proposed when developing a machine learning solution based on a sensor fusion for the operation and maintenance of CANDU reactors:

- If possible, use a dataset that contains readings from a variety of sensors (different part numbers) for machine learning model training.
- Thermal noise is impacted by temperature, therefore using data captured under different temperatures is recommended (if this is applicable such as for systems operating outside).
- When testing, it is recommended to avoid using the cross-validation data set to report testing results. Using a separate test set is recommended.
- If a sensor replacement is required due to failure, re-testing might be required to confirm that the machine learning accuracy is not impacted. Changing the sensor, particularly when the same part number is not available could lead to higher thermal noise which impacts the model performance.
- Model selection shouldn't be only based on the top achieved accuracy. Resilience to thermal noise should be tested according to the expected SNR level. In some cases, it will be hard to test or know the SNR level of the sensor. However, testing the system in an environment that mimics the production environment could be sufficient. If the SNR is known, simulation via noise modeling is recommended.
- During procurement, if sensors' thermal noise levels are known. It is recommended to buy high-quality sensors which are less susceptible to the issue.

## 6. Conclusion

This paper discussed the issue of thermal noise when building a machine learning solution based on sensor fusion for the operation and maintenance of CANDU reactors. Background on the issue and how it can affect sensors reading was presented. Previously published CANDU applications where sensors such as accelerometers, magnetometers, and gyroscopes were included. The paper used an open dataset for a biomedical engineering problem to present how the issue can affect similar future machine learning models which can be built for the operation and maintenance of CANDU reactors. The paper was concluded with a list of recommendations that can help in testing and mitigating the issue.